\documentclass[footinbib,fleqn,aps,prb,twocolumn,superscriptaddress,reprint]{revtex4-2}
\usepackage{xr}
\usepackage[]{amsmath}
\usepackage{amssymb}
\usepackage[dvips]{graphicx}
\usepackage{color}
\usepackage{tabularx}
\usepackage{mathtools}
\usepackage{algpseudocode}
\usepackage{enumitem}
\usepackage{multirow}
\usepackage{epstopdf}  
\usepackage{adjustbox}
\usepackage{bm}
\usepackage{float}
\usepackage{siunitx}
\usepackage{chemformula}
\usepackage{gensymb}
\usepackage[T1]{fontenc}
\usepackage[utf8]{inputenc}

\usepackage{makecell}
\usepackage{hyperref}
\usepackage[version=3]{mhchem}
\usepackage{tikz}
\usetikzlibrary{shapes,snakes}
\usetikzlibrary{arrows}
\usepackage{threeparttable}
\usepackage{natbib}

\newcommand{\Tc}{\textit{T}$_c$}
\newcommand{\WN}{cm$^{-1}$}

\begin{document}

\title{High-pressure stabilization of \ch{Mg2IrH7}: Structural proximity to high-\Tc\ superconductivity}

\author{Shubham Sinha}
\email{ssinha@carnegiescience.edu}
\affiliation{Earth and Planets Laboratory, Carnegie Institution for Science, 5241 Broad Branch Road NW, Washington, DC 20015, USA}

\author{Wencheng Lu}
\email{wlu@carnegiescience.edu}
\affiliation{Earth and Planets Laboratory, Carnegie Institution for Science, 5241 Broad Branch Road NW, Washington, DC 20015, USA}

\author{Mads F. Hansen}
\affiliation{Earth and Planets Laboratory, Carnegie Institution for Science, 5241 Broad Branch Road NW, Washington, DC 20015, USA}
\affiliation{Cavendish Laboratory, University of Cambridge, JJ Thomson Avenue,
Cambridge, CB3 0HE, United Kingdom}

\author{Michael J. Hutcheon}
\affiliation{Theory of Condensed Matter Group, Cavendish Laboratory, J. J. Thomson Avenue, Cambridge CB3 0HE, United Kingdom}

\author{Trevor W. Bontke}
\affiliation{Department of Physics and Texas Center for Superconductivity at the University of Houston, Houston, TX 77204, USA}

\author{Lewis J. Conway}
\affiliation{Department of Materials Science and Metallurgy, University of Cambridge, 27 Charles Babbage Road, Cambridge, CB3 0FS, UK}
\affiliation{Advanced Institute for Materials Research, Tohoku University, Sendai, 980-8577, Japan}

\author{Kapildeb Dolui}
\affiliation{Department of Materials Science and Metallurgy, University of Cambridge, 27 Charles Babbage Road, Cambridge, CB3 0FS, UK}

\author{Chris J. Pickard}
\affiliation{Department of Materials Science and Metallurgy, University of Cambridge, 27 Charles Babbage Road, Cambridge, CB3 0FS, UK}
\affiliation{Advanced Institute for Materials Research, Tohoku University, Sendai, 980-8577, Japan}

\author{Christoph Heil}
\affiliation{Institute of Theoretical and Computational Physics, Graz University of Technology, NAWI Graz, 8010 Graz, Austria}

\author{Piotr A. Gu\'{n}ka}
 \affiliation{Faculty of Chemistry, Warsaw University of Technology, Noakowskiego 3, 00-664 Warszawa, Poland}
 
\author{Stella Chariton}
\affiliation{Center for Advanced Radiation Sources, The University of Chicago, Lemont, Illinois 60439, United States}

\author{Vitali Prakapenka}
\affiliation{Center for Advanced Radiation Sources, The University of Chicago, Lemont, Illinois 60439, United States}

\author{Liangzi Deng}
\affiliation{Department of Physics and Texas Center for Superconductivity at the University of Houston, Houston, TX 77204, USA}

\author{Ching-Wu Chu}
\affiliation{Department of Physics and Texas Center for Superconductivity at the University of Houston, Houston, TX 77204, USA}

\author{Matthew N. Julian}
\affiliation{Enterprise Science Fund, Intellectual Ventures, 3150 139th Ave SE, Bellevue, WA, 98005, USA}

\author{Rohit P. Prasankumar}
\affiliation{Enterprise Science Fund, Intellectual Ventures, 3150 139th Ave SE, Bellevue, WA, 98005, USA}

\author{Timothy A. Strobel}
\email{tstrobel@carnegiescience.edu}
\affiliation{Earth and Planets Laboratory, Carnegie Institution for Science, 5241 Broad Branch Road NW, Washington, DC 20015, USA}

\date{\today}
	
\begin{abstract}
\ch{Mg2IrH6} is a metastable complex metal hydride with a predicted superconducting transition temperature as high as \SI{170}{K} at ambient pressure. Following the synthesis of isomorphic, insulating \ch{Mg2IrH5} at low pressure, higher-pressure studies were conducted to investigate the phase behavior and compound formation in this system. 
X-ray diffraction and Raman spectroscopic measurements indicate that cubic \ch{Mg2IrH7} is stabilized above ca. 40 GPa and coexists with a related hexagonal hydride with likely composition near \ch{Mg2IrH5}. Electrical transport measurements show that cubic \ch{Mg2IrH7} is insulating, in agreement with \textit{ab initio} predictions, and persists during room-temperature decompression until $\sim$\SI{20}{GPa} before reverting back to cubic \ch{Mg2IrH5}. The experimental results confirm ground-state structure predictions in the Mg--Ir--H system, and the formation of two nearly identical phases with surrounding compositions opens new opportunities to access superconducting \ch{Mg2IrH6} through non-equilibrium processing pathways. 

\end{abstract}
\maketitle

\section{Introduction}
Hydrides have helped achieve the long-sought goal of obtaining superconductivity near room temperature~\cite{drozdov2015conventional, somayazulu2019evidence, kong2021superconductivity, Peng2017, ma2022high, semenok2025ternary, song2025}. 
Nevertheless, these materials exhibit such behavior at extremely high pressures, often exceeding hundreds of gigapascals. 
The achievement of superconductivity at such high temperatures represents an important milestone in physics research, however the extreme pressures necessary place major limitations on potential future applications. 
Thus, novel strategies to stabilize these materials or create lower-pressure alternatives is an active research goal.
Lower-pressure stabilization has been realized in some superconducting hydrides such as \ch{LaBeH8} and \ch{LaB2H8}, as well as other classes of superconducting materials~\cite{Chu2022,Deng2025,song2023stoichiometric,song2024superconductivity}.

Recent calculations have proposed conventional high-\Tc\ superconductivity at ambient pressure in several complex metal hydride phases of the form A$_x$MH$_y$ (where A is an alkali, alkaline-earth, or selected trivalent metal, and M is a transition metal, ~\cite{sanna2024prediction,dolui2024feasible,zheng2024prediction, Li2025}), as well as perovskite hydrides such as \ch{AMH3}~\cite{Li2024,Dangic2025}.
Some of these proposed materials are predicted to exhibit \Tc\ well above \SI{100}{K}.
Indeed, prior experimental studies on complex hydrides such as compressed \ch{BaReH9} and \ch{Li5MoH11}~\cite{muramatsu2015metallization, meng2019superconductivity}, as well as low-pressure \ch{Mg4Pt3H6}~\cite{wlu2025} show the possibility of metallic phases that can support superconductivity.

\begin{figure*}
    \centering
    \includegraphics[width=0.9\linewidth]{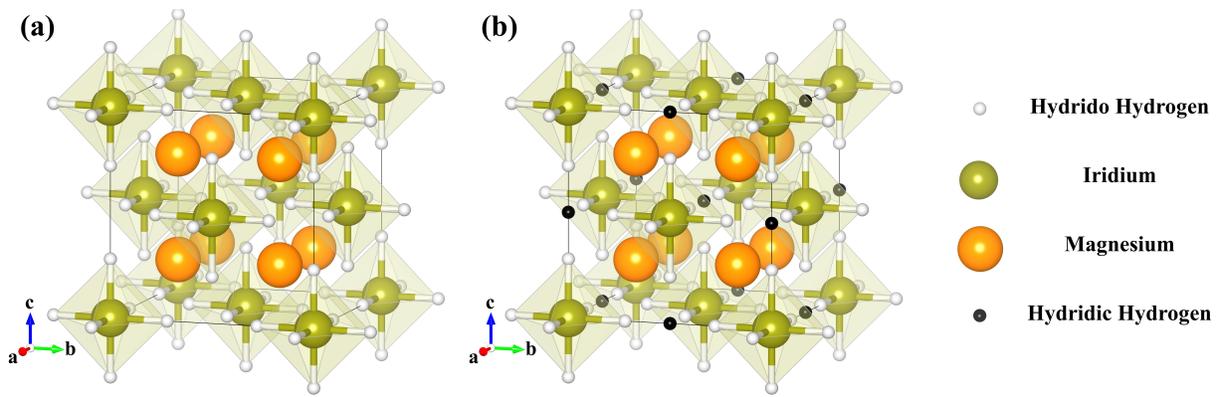}
    \caption{The crystal structures of (a) FCC Mg$_2$IrH$_6$ and (b) FCC Mg$_2$IrH$_7$. The structures are distinguished by an additional interstitial H per formula unit, shown in black.}
    \label{fig:Fig1}
\end{figure*}

Among the proposed superconducting complex hydrides, \ch{Mg2IrH6} (Fig.~\ref{fig:Fig1}) is a promising candidate with predictions of \Tc\ up to \SI{170}{K}~\cite{dolui2024feasible}. 
This phase is predicted to be dynamically stable at ambient pressure and thermodynamically metastable by $\sim$60 meV/atom. Metallic \ch{Mg2IrH6} is comprised of octahedral IrH$_6$ hydrido complexes (formally [IrH$_6$]$^{4-}$), surrounded by Mg$^{2+}$ cations~\cite{dolui2024feasible,Hansen2024}. High-\Tc~superconductivity arises from a significant contribution of H-1$s$ states at a Van-Hove-like singularity near the Fermi level, allowing high-frequency phonons to couple with Fermi-surface electrons~\cite{sanna2024prediction,dolui2024feasible,Xiaoyu2024}.

Recent experimental investigations in the Mg--Ir--H system produced the compound \ch{Mg2IrH5} at pressures between \SI{150}{bar} and $\sim$\SI{30}{GPa}~\cite{Hansen2024}. 
The FCC structure of \ch{Mg2IrH5} is isomorphic with \ch{Mg2IrH6}, but with an octahedral hydrogen site occupancy of $\frac{5}{6}$, resulting in disordered [IrH$_5$]$^{4-}$ complexes with a characteristic Ir--H Raman stretching mode near \SI{2100}{\per\cm} at ambient pressure.
The two Mg$^{2+}$ cations per formula unit compensate the charge of the [IrH$_5$]$^{4-}$ complex anions, yielding a charge-balanced insulator. 
\ch{Mg2IrH5} is one hydrogen vacancy away from superconducting \ch{Mg2IrH6}, which might be achieved via low-barrier hydrogen insertion~\cite{Hansen2024}.

From another side, \ch{Mg2IrH6} might be obtained by hydrogen removal from FCC \ch{Mg2IrH7}. \ch{Mg2IrH7} was predicted to be thermodynamically stable above $\sim$\SI{15}{GPa}, and the removal of a single interstitial hydridic hydrogen---the distinguishing structural feature between \ch{Mg2IrH6}---could enable the formation of the superconducting phase~\cite{dolui2024feasible}. While \ch{Mg2IrH7} was predicted to be thermodynamically stable near \SI{15}{GPa}, it was not observed experimentally to pressures around \SI{30}{GPa}~\cite{Hansen2024}. Configurational entropy of \ch{Mg2IrH5}, originating from the disordered [IrH$_5$]$^{4-}$ complexes, contributes to the stability of this phase and may alter the predicted ground-state phases at finite temperature~\cite{dolui2024feasible}. Given that static calculations show the stability of \ch{Mg2IrH5} decreases above \SI{15}{GPa} while \ch{Mg2IrH7} remains on the calculated convex hull, it is plausible that \ch{Mg2IrH7} becomes stabilized on basis of free energy at pressures high enough to overcome the configurational entropy contribution of \ch{Mg2IrH5}~\cite{Hansen2024, dolui2024feasible}.         

In this study, we have conducted experiments on \ch{Mg2IrH5} to higher pressures near \SI{40}{GPa}, finding that it transforms into FCC \ch{Mg2IrH7} in the presence of excess hydrogen. \ch{Mg2IrH7} is clearly distinguished from \ch{Mg2IrH5} by a strong [IrH$_6$]$^{3-}$ Raman stretching mode near $\sim$1500 cm$^{-1}$, softened by roughly 800 cm$^{-1}$ from the [IrH$_5$]$^{4-}$ units within \ch{Mg2IrH5} at \SI{40}{GPa}, as well as an expanded FCC lattice volume. FCC \ch{Mg2IrH7} typically coexists with another energetically competitive hexagonal hydride with composition likely near \ch{Mg2IrH5}. \ch{Mg2IrH7} persists during decompression until $\sim$\SI{20}{GPa}, before reverting back to \ch{Mg2IrH5}. The observation of FCC \ch{Mg2IrH7}, which is nearly equivalent to \ch{Mg2IrH6}, presents additional opportunities to realize high-\Tc\ hydride superconductivity at reduced-pressure conditions.

\section{Methods}

\textit{Experimental synthesis}. The \ch{Mg2IrH5} samples used as precursors in this study were synthesized by mixing Mg (Sigma Aldrich 99.5\%) and Ir (Sigma Aldrich 99.9\%) in a 2:1 molar ratio which were then heated inside a stainless steel autoclave-type reactor at 450 \degree C under a hydrogen (UHP Airgas) pressure of $\sim$\SI{150}{bar} for two weeks. 
The recovered precursor consisted of FCC \ch{Mg2IrH5} with small impurities of unreacted FCC Ir and a $Pm\bar{3}m$ MgIrH$_x$ compound. Additional synthesis details can be found in reference~\cite{Hansen2024}.

High-pressure experiments were conducted using symmetric and BX90 diamond anvil cells (DACs) equipped with \SI{300}{\micro\m} culet anvils. Rhenium gaskets were indented to \SI{40}{\micro\m} and a 150--\SI{200}{\micro\m} hole was laser drilled for the sample chamber. 
Thin pellets of the \ch{Mg2IrH5} precursor were created by pressing powder between two \SI{1}{mm} culet anvils, small pieces from which were then loaded into the sample chamber and filled with either fluid H$_2$ at $\sim$1.5 kbar, or ammonia borane, which served as a source for excess hydrogen. Sample pressures were estimated at room temperature using the diamond Raman edge or from a ruby chip placed in the sample chamber~\cite{ruby, akahama2010pressure}.

Samples were compressed to target pressures near \SI{40}{GPa}, and one- or two-sided laser heating was performed using a 1064 nm Nd:YAG laser~\cite{Prakapenka2008}. It was found that low laser powers near the laser-coupling threshold ($\sim$800 K) were sufficient to induce a structural transformation, and temperature estimations based on thermal emission were not viable in most low-power experiments. Additional heating at higher temperatures resulted in the formation of a more complex hexagonal structure. Several heating tests were also performed using the $\sim$10 $\mu$m of a \ch{CO2} laser, which was found to produce equivalent results.

\textit{Raman spectroscopy}. Unpolarized Raman spectra were measured using a 532 nm laser for excitation focused through a 20$\times$ long-working-distance objective lens at a power of $\sim$ 5 mW, combined with a Princeton instruments SP2750 spectrograph with a 750 mm focal length.

\textit{X-ray diffraction}. Angular-dispersive, monochromatic X-ray diffraction measurements were carried out in transmission geometry at 13-IDD (GSECARS) at the Advanced Photon Source using a wavelength of 0.3344 \AA~using a Pilatus3 X CdTe 1M detector. The sample-to-detector distance and detector geometry were calibrated using a \ch{LaB6} standard, in conjunction with the Dioptas software \cite{prescher2015dioptas}, which was also used for data reduction. Preliminary diffraction measurements were also performed at beamline 12.2.2 \cite{Kunz:xd5008} at the Advanced Light Source using a wavelength of 0.4959 \AA~using a Pilatus Si detector. For \textit{in situ} diffraction measurements, pressure determination was established using the equation of state (EOS) of Ir \cite{Monteseguro2019} from small regions of unreacted sample near the laser heating spot, in addition to diamond Raman measurements~\cite{akahama2010pressure}. To establish lattice parameters, Le Bail refinements of powder XRD data were performed using GSAS-II~\cite{toby2013gsas}.

Crystalline diffraction spots were observed from a sample heated near \SI{1500}{K}, which enabled multigrain analysis using a synchrotron beam size of $\sim$2$\times$\SI{2}{\micro\m}. Frames were recorded between an angular range of $\omega = -32^\circ$ to +$31^\circ$ in 0.5$^\circ$ steps with an exposure time of \SI{1}{s} per frame. Reflections were harvested using CrysAlisPRO, assigned to individual grains with DAFi~\cite{Aslandukov2022}, and subsequently indexed and integrated using CrysAlisPRO software suite~\cite{noauthor_crysalispro_2023}. Crystal structures were solved with SHELXT 2018/2 and refinement was attempted using SHELXL 2018/3, invoked from within the Olex2 suite~\cite{sheldrick_crystal_2015, sheldrick_shelxt_2015, dolomanov_olex2_2009}.

\textit{Electrical transport measurements}. Non-magnetic DACs compatible with Quantum Design Physical Property Measurement System (PPMS model 6000) were utilized for electrical transport measurements with the maximum excitation current set to \SI{1}{mA}. Two flat, \SI{300}{\micro\m} culets were used with a Re gasket, which was indented to a thickness of $\sim$\SI{40}{\micro\m} and laser-drilled with a $\sim$\SI{280}{\micro\m} hole. A layer of cubic boron nitride (cBN)--epoxy composite was compressed for an insulation layer on one side of the gasket and then re-drilled with a $\sim$\SI{130}{\micro\m} hole for the sample chamber. The chamber was next filled with ammonia borane for the pressure medium and hydrogen source, in which a thin pellet of \ch{Mg2IrH5} was embedded. Pt electrodes cut from \SI{12}{\micro\m}-thick foil were then placed onto the sample pellet in van der Pauw configuration for four probe measurement ~\cite{VanderPauw1991}.

\textit{DFT calculations}. Geometry optimizations and all subsequent calculations were performed using \textsc{castep} \cite{clark2005first} with the Perdew-Burke-Ernzerhof generalized gradient approximation for solids (PBEsol) \cite{PhysRevLett.100.136406}, a \SI{600}{eV} plane-wave cutoff, a $k$-point spacing of 2$\pi\times$0.03 \AA$^{-1}$, and default \textsc{castep} norm-conserving  pseudopotentials (NCP). Structures were converged to give forces and stresses to within \SI{0.05}{eV}$\cdot$\AA$^{-1}$ and \SI{0.01}{GPa}. Gamma-point phonon and Raman intensity calculations were performed using density functional perturbation theory implemented in \textsc{castep}~\cite{Refson2006} and \textsc{quantum espresso}~\cite{giannozzi2009quantum}. \textsc{quantum espresso} calculations were performed using the Perdew-Burke-Ernzerhof (PBE) functional \cite{perdew1996generalized} and v1.5 of the GBRV pseudopotentials \cite{Garrity2014}, with comparable parameters and tolerances to the \textsc{castep} calculations.

\section{Results and discussion}

\begin{figure}[t!]
\begin{center}
\includegraphics[width=0.48\textwidth]{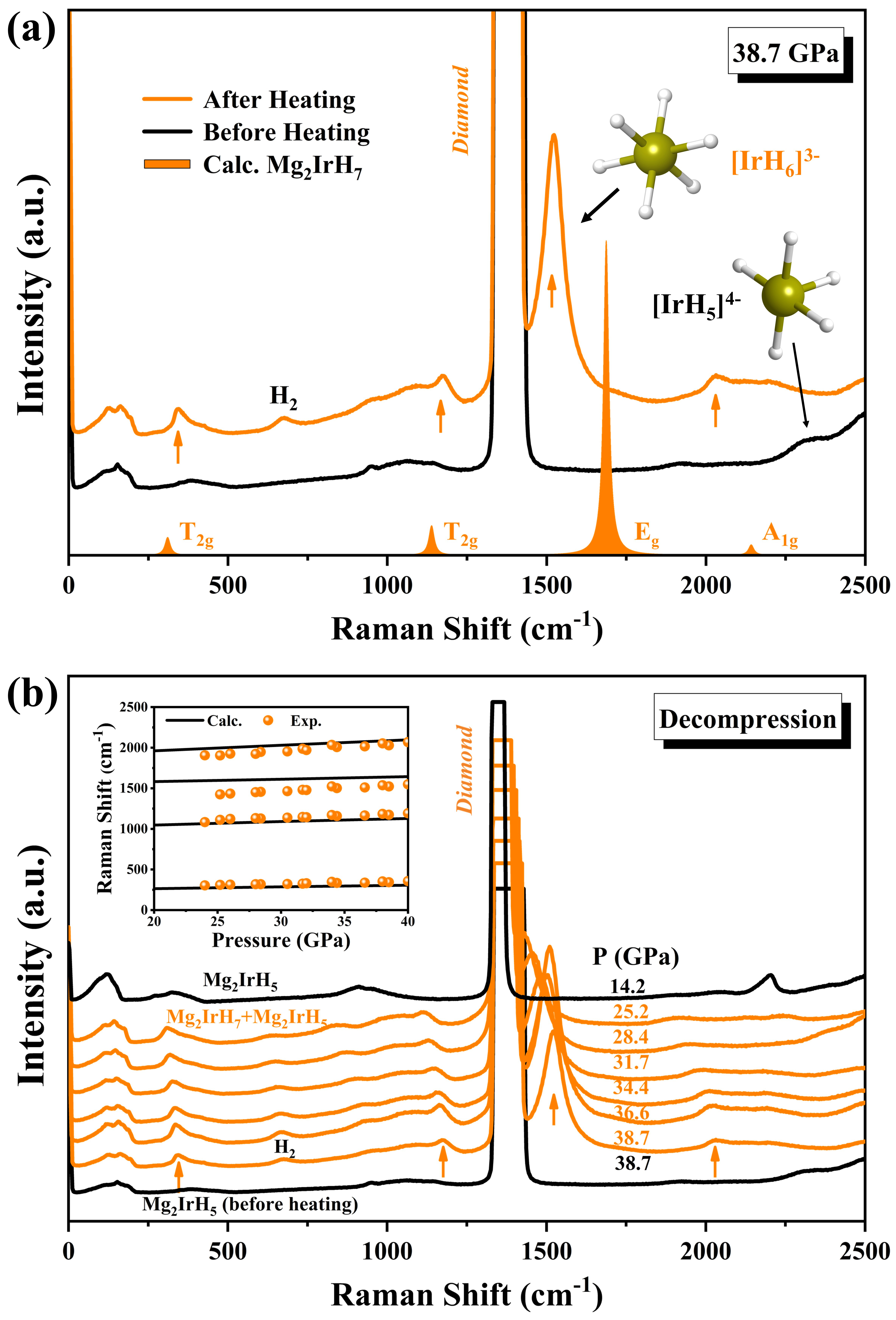}\\[5pt] 
\caption{(a) Experimental Raman spectra of \ch{Mg2IrH5} before and after heating at \SI{38.7}{GPa} in ammonia borane. The calculated spectrum of \ch{Mg2IrH7} is shown for comparison using Lorentzian peak shapes with arbitrary widths. Orange arrows highlight new peaks attributed to [IrH6]$_{3-}$ observed after heating.(b) Raman spectra of a heated sample obtained at room temperature while decompressing from 38.7 GPa. The inset compares experimental peak positions for \ch{Mg2IrH7} (points) with DFT-calculated (QE-PBE) frequencies (lines).}
\label{fig:Fig2}
\vspace{-0.5cm} 
\end{center}
\end{figure}

The Raman spectrum of \ch{Mg2IrH5} is characterized by an Ir--H stretching mode at $\sim$\SI{2100}{\per\centi\metre}, Ir--H bending modes near \SI{800}{\per\centi\metre}, and lattice phonons at lower frequency~\cite{Hansen2024}. When compressed to \SI{38.7}{GPa} at room temperature, these modes broaden significantly and the stretching vibration of [IrH$_5$]$^{4-}$ is detectable only as a shoulder near $\sim$\SI{2300}{\per\centi\metre} at the onset of the second-order diamond Raman peak (Fig.~\ref{fig:Fig2}a). Gentle laser heating ($<$\SI{1000}{K}) of compressed \ch{Mg2IrH5} with a hydrogen source (either \ch{H2} or ammonia borane) produced marked changes in the Raman spectrum. After heating at \SI{38.7}{GPa}, a very intense peak appears at \SI{1523}{\per\centi\metre}, as well as three additional weaker, yet distinct peaks at 343, 1174 and \SI{2029}{\per\centi\metre} (Fig.~\ref{fig:Fig2}a). 

\begin{figure}[b!]
\begin{center}
\includegraphics[width=0.48\textwidth]{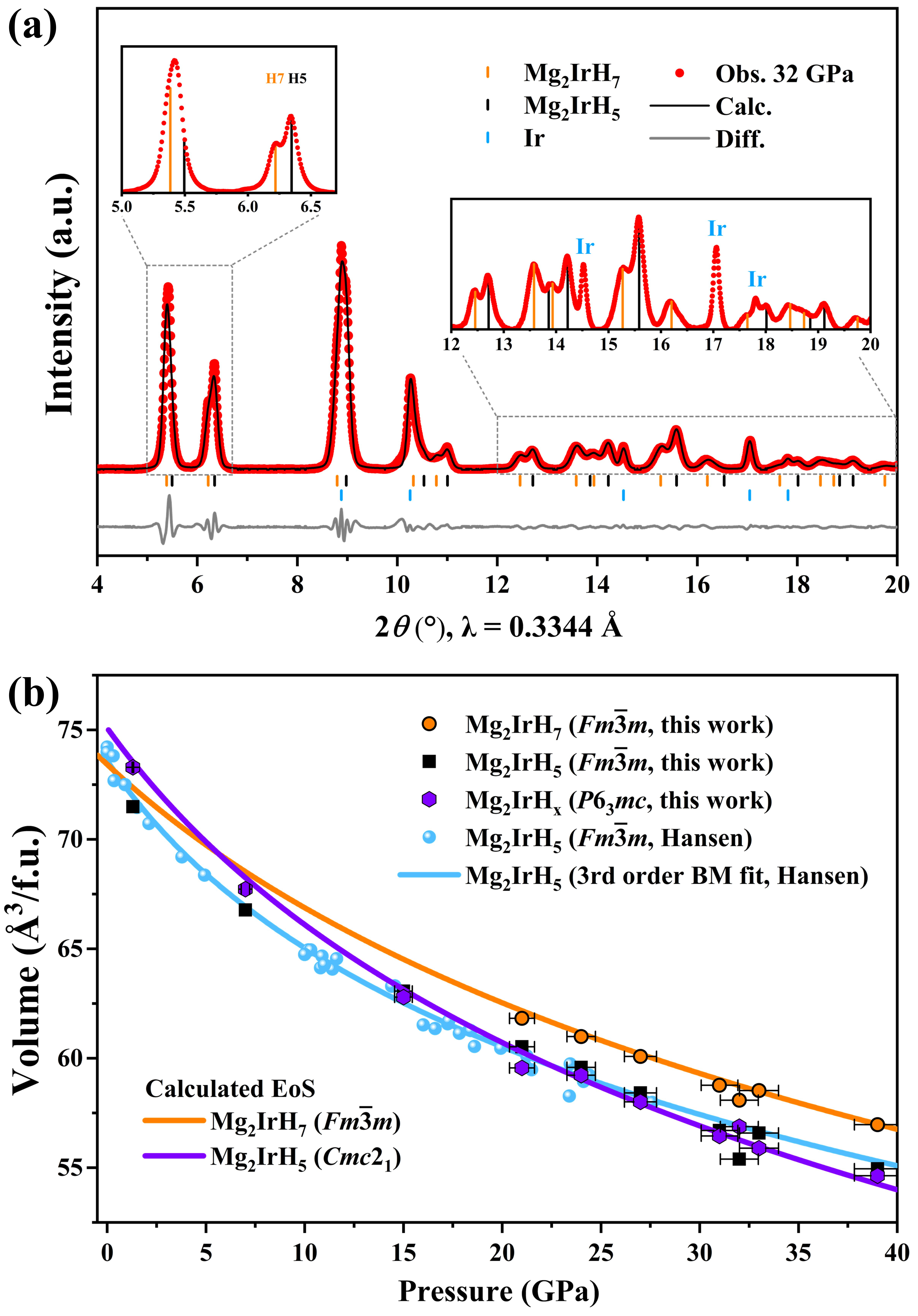}\\[5pt] 
\caption{(a) Representative XRD pattern showing the formation of an expanded FCC phase after heating \ch{Mg2IrH5} in hydrogen. The sample pressure dropped from \SI{40}{GPa} to \SI{32}{GPa} after heating. The red points correspond to the experimental data, while the black and gray curves represent the refinement
and residual, respectively. Orange, black, and blue ticks indicate the calculated peak positions for \ch{Mg2IrH7}, \ch{Mg2IrH5} and Ir. (b) Volume per formula unit plotted as a function of pressure for \ch{Mg2IrH5},~\cite{Hansen2024} \ch{Mg2IrH7}, and hexagonal Mg$_2$IrH$_x$. The orange and purple lines show the DFT-calculated equations of state. Ordered $Cmc2_1$ \ch{Mg2IrH5} is used to approximate the disordered hexagonal variation.}
\label{fig:Fig3}
\vspace{-0.5cm} 
\end{center}
\end{figure}

The strong Raman intensity of the peak at \SI{1523}{\per\centi\metre} is suggestive of Ir--H stretching, and the frequency is consistent with the formation of [IrH$_6$]$^{3-}$ units based on comparison with other hydrido complexes MH$_6$ octahedral units (e.g., \ch{Mg2FeH6}, \ch{Ca2RuH6}, \ch{Sr2RuH6}, \ch{Eu2RuH6}, \ch{Mg2OsH6}, \ch{Na3CoH6}, and \ch{Rb2PtH6})~\cite{1997_Parker,parker2003inelastic,hagemann2002raman,parker2010spectroscopy}. The nearly \SI{800}{\per\centi\metre} frequency shift from [IrH$_5$]$^{4-}$ is diagnostic of charge redistribution and strongly suggestive of a compound containing the [IrH$_6$]$^{3-}$ complex. Note that the DFT-relaxed Ir--H bond distance for [IrH$_6$]$^{3-}$ within \ch{Mg2IrH7} / \ch{Mg2IrH6} is \SI{1.71}{\angstrom}, significantly elongated from \SI{1.66}{\angstrom} for [IrH$_5$]$^{4-}$ within \ch{Mg2IrH5}. 

To help understand the structural transformation observed upon heating, we calculated the Raman frequencies and intensities for \ch{Mg2IrH7}, which was predicted to be the stable high-pressure phase at this Mg:Ir ratio~\cite{dolui2024feasible}. The $Fm\bar3m$ structure possesses four Raman active phonons calculated at 310 ($T_{2g}$), 1139 ($T_{2g}$), 1687 ($E_{g}$) and 2142 ($A_{1g}$) \WN\ at \SI{40}{GPa}. We note that \ch{Mg2IrH6} in the same space group possesses the same four Raman phonons at similar frequencies, but the second $T_{2g}$ bending mode is softened by $\sim$\SI{200}{\per\centi\metre} in the absence of the interstitial hydrogen. The calculated frequencies of the four Raman modes for \ch{Mg2IrH7} are in good agreement with the four peaks observed experimentally, noting that the calculated frequencies under the static and harmonic approximations vary by up to $\sim$\SI{100}{\per\centi\metre} depending on the specific DFT functional (e.g., PBE vs. BLYP) and code (e.g., \textsc{castep} vs. \textsc{quantum espresso}).

Observed across multiple independent experiments, the Raman frequencies of the four new peaks systematically decrease upon decompression and, below $\sim$\SI{20}{GPa}, the system reproducibly reverts to \ch{Mg2IrH5} (Fig.~\ref{fig:Fig2}(b) and Fig. S1). This reverse transformation is evidenced by the reappearance of the diagnostic [IrH$_5$]$^{4-}$ stretching band near \SI{2200}{\per\centi\metre}. The reversion to \ch{Mg2IrH5} during decompression at room temperature is indicative of a low energy barrier and supportive of the \ch{Mg2IrH7} Raman interpretation given the strong structural similarity between the two phases. The pressure dependence of the four observed Raman peaks, which track calculated mode frequencies (Fig.~\ref{fig:Fig2}(b), inset), further supports the formation of \ch{Mg2IrH7}.

To elucidate the structure of the new compound formed near \SI{40}{GPa}, we performed \textit{in situ} synchrotron XRD measurements on samples loaded in hydrogen. Before heating at 38 GPa, samples are well characterized by FCC \ch{Mg2IrH5} with broadened Bragg peaks, see Fig. S2 in Supplementary Material. Some regions of the sample contained small amount of metallic Ir or MgIrH$_x$ impurities, also present in the starting precursor. The lattice parameter (see more structural details in Table S1) of $a=$ \SI{6.032(5)}{\angstrom} is in good agreement with extrapolations from the previous report \cite{Hansen2024}. 

\begin{figure*}
    \centering
    \includegraphics[width=1\linewidth]{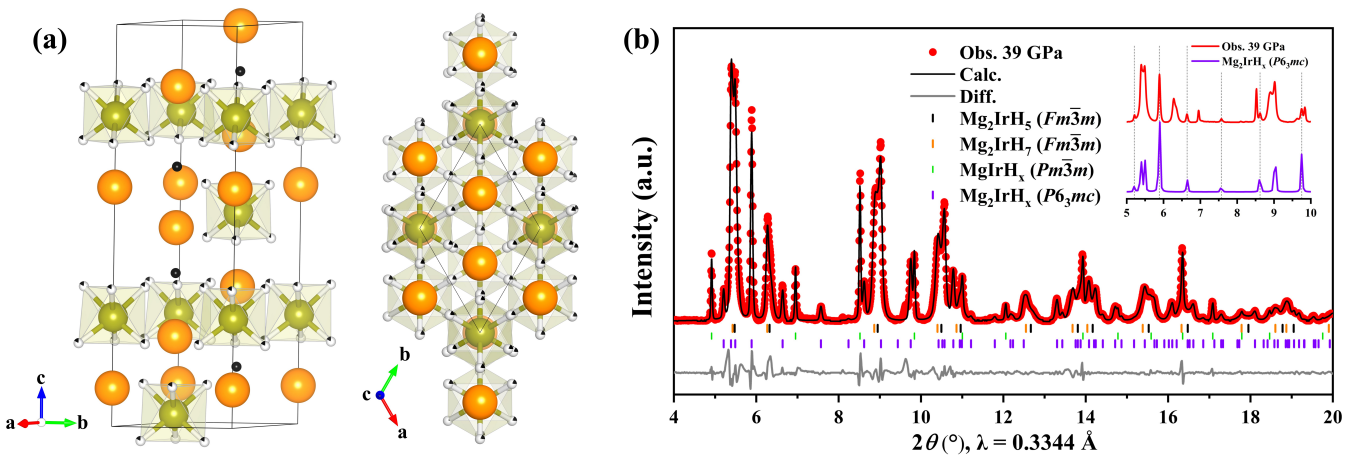}
    \caption{(a) Crystal structure of $P6_3mc$ Mg$_2$IrH$_x$. The calculated $P6_3mc$ \ch{Mg2IrH7} structure possesses [IrH$_6$]$^{3-}$ units and intersitial H atoms (black spheres), like $Fm\bar3m$ \ch{Mg2IrH7}. In space group $P6_3mc$, \ch{Mg2IrH5} contains disordered [IrH$_5$]$^{4-}$ units (i.e., $\frac{5}{6}$ H occupancy represented by pie-chart spheres) with no interstitial hydrogen and can be described as an ordered orthorhombic $Cmc$2$_1$ structure in calculations.  (b) Representative powder XRD pattern obtained after laser heating  the samples to \SI{1500}{K} and corresponding Le Bail refinement. The red points correspond to the experimental data, while the black and gray curves represent the refinement and residual, respectively. Orange, black, green, and purple ticks indicate the calculated peak positions for $Fm\bar3m$ \ch{Mg2IrH7}, $Fm\bar3m$ \ch{Mg2IrH5}, $Pm\bar3m$ MgIrH$_x$, and $P6_3mc$ Mg$_2$IrH$_x$. Inset: comparison of experimental data and calculated powder intensity profile for the $P6_3mc$ structure. Dashed lines mark select predicted Bragg positions for the $P6_3mc$ structure that do not overlap with coexisting phases.}
    \label{fig:Fig4}
\end{figure*}

Immediately upon heating samples at the lowest laser powers possible (i.e., $T < \SI{1000}{\kelvin}$), we observed the distinct splitting of the FCC peaks (Fig.~\ref{fig:Fig3}(a)), indicating the formation of a second, expanded FCC phase with {$a=$ \SI{6.179}{\angstrom}}. The presence of these new FCC peaks are coincident with the formation of the new Raman peaks described above, particularly the diagnostic peak near \SI{1500}{\per\centi\metre} attributed to [IrH$_6$]$^{3-}$ stretching. The observed volume expansion from \ch{Mg2IrH5} to \ch{Mg2IrH7} is consistent with DFT calculations, which indicate a 3.9\% volume increase at \SI{40}{GPa}.

The expanded FCC lattice persists during decompression, tracking the DFT-calculated equation of state for \ch{Mg2IrH7} (Fig.~\ref{fig:Fig3}(b)), until approximately \SI{20}{GPa}. Below this pressure, the expanded lattice vanishes and only \ch{Mg2IrH5} was observed. The correspondence between XRD and Raman measurements allow us to confidently attribute the observed Raman modes to the expanded FCC phase, and agreement with DFT calculations strongly supports the assignment of this phase to FCC \ch{Mg2IrH7}.

Additional heating of samples (i.e., $T > \SI{1000}{\kelvin}$) resulted in the formation of a more complex phase with many diffraction peaks (Fig. \ref{fig:Fig4}(b) and Fig. S3). The appearance of this phase was observed in some locations using the lowest heating powers and its formation is difficult to avoid, though it is favored at higher temperatures. Using single-crystal diffraction techniques with multigrain analysis on a sample heated near \SI{1500}{K}, we were able to index this phase to a hexagonal lattice with $a=$ \SI{4.31}{\angstrom} and $c=$ \SI{14.2}{\angstrom} at 32 GPa, reproduced for a total of 22 grains at three sample positions. A tentative structural solution was obtained via intrinsic phasing, however subsequent refinements failed to converge with high reliability (see Table S2). Nevertheless, the analysis provides useful insights into the metal substructure, which apparently maintains the starting \ch{Mg2Ir} composition ($Z=4$) in the likely space group $P$6$_3$$mc$ with three Mg atoms located on the 2$b$ positions ($z$ $\approx$ 0.22, 0.36, and 0.38), one Mg atom on the 2$a$ position ($z$ $\approx$ 0.095), and two Ir atoms on the 2$a$ and 2$b$ positions ($z$ $\approx$ 0.29 and 0.54, respectively). The metal substructure contains several voids large enough to host hydrogen atoms.

While we were unable to explicitly solve the crystal structure and hydrogen content of this hexagonal phase, we performed constrained AIRSS runs, informed by the hexagonal experimental lattice, finding low-energy candidate structures that match the experimental observations. Interestingly, two of the compounds have the compositions \ch{Mg2IrH6} and \ch{Mg2IrH7} with space group $P$6$_3$$mc$ and atomic metal positions similar to those produced by experiments (see Supplemental Information). At \SI{30}{GPa}, these phases are located 30--\SI{40}{meV/atom} above the convex hull and may be viewed as hexagonal transformations of the FCC phases with a doubling of the $c$-axis. While this hexagonal variation was not predicted for \ch{Mg2IrH5}, an orthorhombic $Cmc$2$_1$ phase with this composition was found with similar lattice vectors (see Table S3) following $b_{ortho} \approx b_{hex} \sqrt{3}$. Similar to the case of FCC \ch{Mg2IrH5}---where an ordered tetragonal structure was calculated to approximate the disordered FCC compound observed experimentally---this ordered $Cmc$2$_1$ version of \ch{Mg2IrH5} may represent a disordered hexagonal structure that is favored at finite temperature. Based on observed formula unit volumes of the hexagonal phase with pressure, the \ch{Mg2IrH5} composition is in good agreement with calculations (Fig. \ref{fig:Fig3}(b)). In this case, the hexagonal phase can represent a hexagonal arrangement of the disordered FCC \ch{Mg2IrH5} structure that becomes more favorable around \SI{40}{GPa}, similar to pressure-induced martensitic transformations observed for noble gases~\cite{Kim2006, Rosa2018}.

The observation of this hexagonal structure raises the question of whether the observed Raman modes described above could be attributed to this phase. In addition to the observation of the disappearance of these Raman modes at \SI{20}{GPa}, coincident with the disappearance of the expanded FCC lattice by XRD at the same pressure, this possibility is ruled out by the preservation of the hexagonal phase to lower pressures by XRD. Hexagonal Mg$_2$IrH$_x$ was observed to the lowest measured pressure (\SI{1.3}{GPa}), well beyond the disappearance of the Raman spectrum attributed to cubic \ch{Mg2IrH7}. If the composition of the hexagonal phase is indeed \ch{Mg2IrH5}, as supported by obtained diffraction volumes, it is likely that molecular Raman modes would largely coincide with FCC \ch{Mg2IrH5} given the presence of identical, disordered constituents. 

\begin{figure}[t!]
\begin{center}
\includegraphics[width=0.48\textwidth]{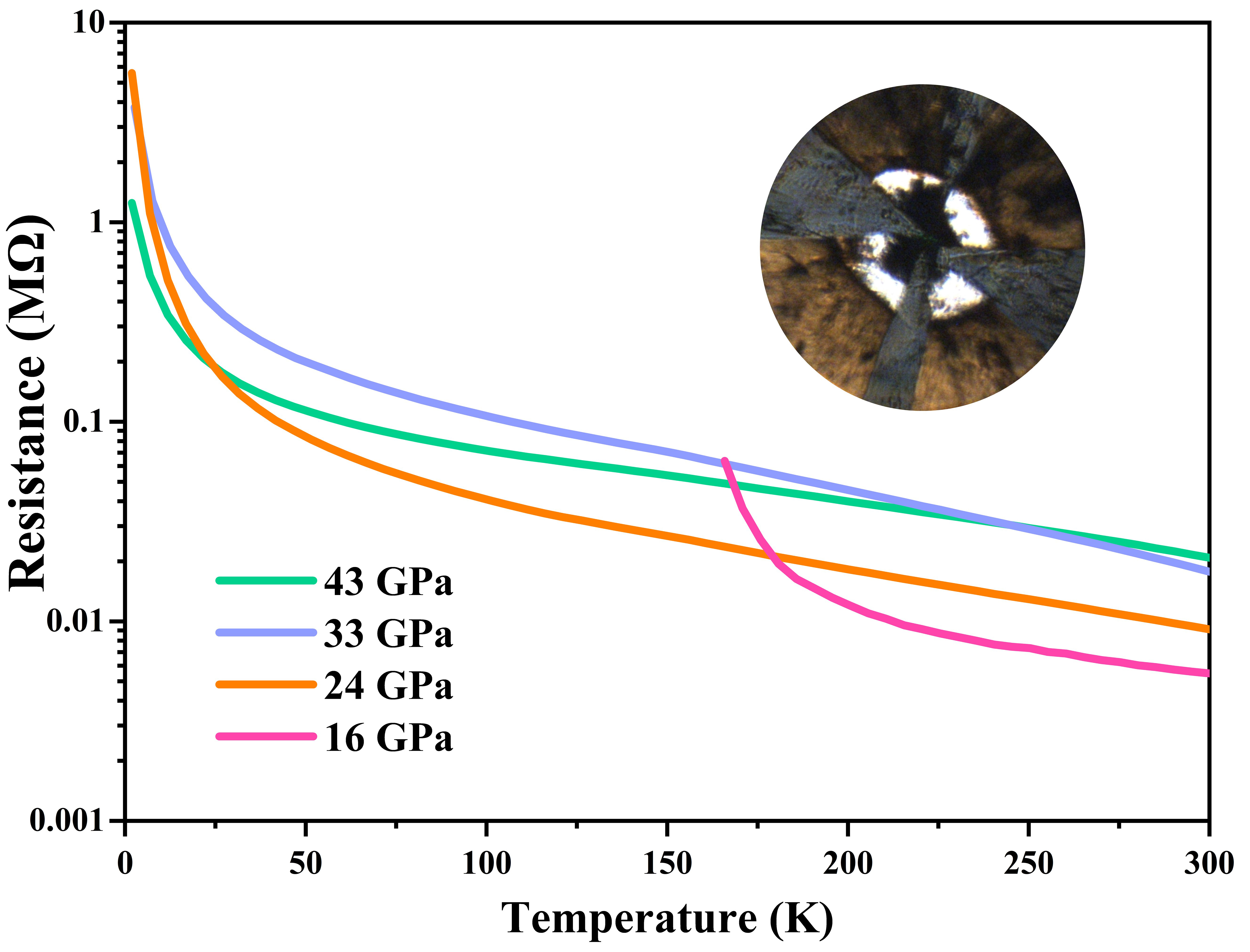}\\
\caption{Temperature-dependent electrical transport measurements performed after the synthesis of FCC \ch{Mg2IrH7} at 43 GPa and at subsequent decompression steps. Inset: \ch{Mg2IrH5} sample pellet with electrical contacts embedded in ammonia borane. The culet diameter is \SI{300}{\micro\m}.}
\label{fig:Fig5}
\vspace{-0.8cm} 
\end{center}
\end{figure}

Finally, the formation of FCC \ch{Mg2IrH7} allows us to begin the exploration of synthetic pathways to superconducting \ch{Mg2IrH6} via hydrogen removal, as initially proposed by Dolui et al.~\cite{dolui2024feasible}. Given the structural similarity with \ch{Mg2IrH6}, diffusion of the non-covalent hydrogen from \ch{Mg2IrH7} represents a potential route to obtain this superconducting metastable phase. This possibility might be realized via the decompression of \ch{Mg2IrH7} given the local, low-energy minimum of \ch{Mg2IrH6} before reversion to \ch{Mg2IrH5}, and would be manifested by a change to metallic conductivity/superconductivity at some intermediate pressure.

To study this possibility, the behavior of \ch{Mg2IrH7} was investigated using electrical transport measurements. A \ch{Mg2IrH5} sample equipped with electrical contacts was heated in ammonia borane near \SI{40}{GPa} and the presence of an expanded FCC lattice was verified by XRD, in addition to the presence of the strong $\sim$1520 \WN\ Raman mode mapped across the sample. Temperature-dependent transport measurements near \SI{40}{GPa} show insulating behavior with resistance reaching the M$\Omega$ level at low temperature (Fig.~\ref{fig:Fig5})---the expected behavior for FCC \ch{Mg2IrH7}, which is predicted to be a charge-balanced insulator (i.e., 2Mg$^{2+}$$\cdot$[IrH$_6$]$^{3-}$$\cdot$H$^-$) with a PBE-DFT-predicted band gap of \SI{2.46}{eV}.

When temperature-dependent transport measurements were conducted on \ch{Mg2IrH7} at pressure steps during decompression, the sample remained insulating with an offset in the resistance magnitude across the $\sim$\SI{20}{GPa} decomposition threshold, presumably related to hydrogen loss and the reversion back to the \ch{Mg2IrH5} structure. Thus, the formation of \ch{Mg2IrH6} does not appear to occur spontaneously at room temperature based on decompression measurements performed across the pressure stability field. Future measurements focused on the kinetics and detailed \ch{Mg2IrH7} \textrightarrow~\ch{Mg2IrH5} transition mechanism could produce valuable insights for the transient stabilization of superconducting \ch{Mg2IrH6}.

While metastable \ch{Mg2IrH6} remains elusive to date, the demonstration of two nearly identical structures is promising for its future synthesis. FCC \ch{Mg2IrH5}, and now, \ch{Mg2IrH7} are available as starting precursors with isostructural metal substructure arrangements, differing only by a single hydrogen atom per formula unit. Non-equilibrium processing methods such as implantation/irradiation, deposition, electro/mechanochemistry, or dynamic P/T cycling might serve as valuable routes to be explored to isolate this exceptional, dynamically stable phase.

\section{Conclusion}
Following prior studies on FCC \ch{Mg2IrH5}, $Fm\bar3m$ \ch{Mg2IrH7} was found to form at higher pressures near 40 GPa. The Raman spectrum of \ch{Mg2IrH7} is characterized by an intense Ir--H stretching band near 1520 \WN, diagnostic of [IrH$_6$]$^{3-}$ complexes, clearly distinguishing this phase from \ch{Mg2IrH5}. Compared with FCC \ch{Mg2IrH5}, \ch{Mg2IrH7} has an expanded lattice and persists to ca. 20 GPa when decompressed at room temperature. In addition to FCC \ch{Mg2IrH7}, another hexagonal Mg$_2$IrH$_x$ hydride was observed upon heating samples near 40 GPa. This structure crystallizes in the probable space group $P6_3mc$ with the likely composition \ch{Mg2IrH5} based on formula-unit volume, and likely represents a hexagonal modification of disordered FCC \ch{Mg2IrH5} with a doubling of the $c$-axis. The thermodynamic stability, spectroscopic behavior and structural observations in this work are all strongly supported by \textit{ab intio} calculations, providing further confidence in experimental interpretations and in using these tools to help guide materials discovery. While superconducting \ch{Mg2IrH6} was not observed as a spontaneous intermediate phase during the decompression of \ch{Mg2IrH7}, the successful formation of \ch{Mg2IrH7} enables access to a structurally related compound that may serve as a useful precursor for future non-equilibrium synthesis approaches.

\begin{acknowledgments}
This work was supported by the Deep Science Fund of Intellectual Ventures. We acknowledge the GeoSoilEnviroCARS (The University of Chicago, Sector 13), Advanced Photon Source, Argonne National Laboratory for provision of synchrotron radiation facilities at beamline. GeoSoilEnviroCARS is supported by the National Science Foundation – Earth Sciences via SEES: Synchrotron Earth and Environmental Science (EAR –2223273). This research used resources of the Advanced Photon Source, a U.S. Department of Energy (DOE) Office of Science User Facility operated for the DOE Office of Science by Argonne National Laboratory under Contract No. DE-AC02-06CH11357. Use of the GSECARS Raman Lab System was supported by the NSF MRI Proposal (EAR-1531583). This research also used resources of the Advanced Light Source, a U.S. DOE Office of Science User Facility under contract no. DE-AC02-05CH11231. The authors thank Bora Kalkan and Martin Kunz for assistance with XRD measurements and Amol Karandikar for assistance with \ch{CO2} laser heating.
\end{acknowledgments}


\bibliography{ref}

\end{document}